\documentstyle[12pt]{article}
\def\thebibliography#1{\section*{References}\list
  {[\arabic{enumi}]}{\settowidth\labelwidth{#1}\leftmargin\labelwidth
    \advance\leftmargin\labelsep
    \usecounter{enumi}}
    \def\newblock{\hskip .11em plus .33em minus .07em}
    \sloppy\clubpenalty4000\widowpenalty4000
    \sfcode`\.=1000\relax}

\def\op#1{\mathop{\fam0 #1}\limits}

\newcommand{\nm}[1]{\mid {#1}\mid}

\newcommand{\beq}{\begin{equation}}
\newcommand{\eeq}{\end{equation}}
\newcommand{\ben}{\begin{eqnarray}}
\newcommand{\een}{\end{eqnarray}}
\newcommand{\be}{\begin{eqnarray*}}
\newcommand{\ee}{\end{eqnarray*}}
\newcommand{\bea}{\begin{eqalph}}
\newcommand{\eea}{\end{eqalph}}
\newcommand{\bR}{{\bf R}}
\newcommand{\bL}{{\bf L}}
\newcommand{\al}{\alpha}
\newcommand{\bt}{\beta}
\newcommand{\la}{\lambda}
\newcommand{\f}{\phi}

\newcommand{\x}{\xi}
\newcommand{\om}{\omega}
\newcommand{\m}{\mu}
\newcommand{\n}{\nu}
\newcommand{\dl}{\delta}
\newcommand{\g}{\gamma}

\newcommand{\si}{\sigma}
\newcommand{\Si}{\Sigma}
\newcommand{\kp}{\kappa}
\newcommand{\ve}{\varepsilon}

\newcommand{\wt}{\widetilde}

\newcommand{\dr}{\partial}
\newcommand{\ap}{\approx}

\newcommand{\ot}{\otimes}

\newcounter{eqalph}
\newcounter{equationa}

\newenvironment{eqalph}{\stepcounter{equation}
\setcounter{equationa}{\value{equation}}
\setcounter{equation}{0}

\begin{eqnarray}}{\end{eqnarray}\setcounter{equation}{\value{equationa}}}

\begin{document}

\hbox{}

\begin{center}

{\large \bf Energy-momentum and gauge conservation laws}
\bigskip

G.Giachetta, L.Mangiarotti,$^1$ and G.Sardanashvily$^2$
\medskip

{\it $^1$Department of Mathematics and Physics, University of Camerino,
Camerino 62032, Italy

$^2$Department of Theoretical Physics, Moscow State University, 117234
Moscow, Russia}
\bigskip
\end{center}

{\small
We treat energy-momentum conservation laws as particular gauge conservation
laws when generators of gauge transformations are
horizontal vector fields on fibre bundles. In particular, the generators
of general covariant transformations are the  canonical horizontal
prolongations of vector fields on a world manifold. This is the case  of the
energy-momentum conservation in gravitation theories. 
We find that, in 
main gravitational models, the corresponding energy-momentum flows reduce to
the generalized Komar superpotential.
We show that
the superpotential form of a conserved flow is the common property of gauge
conservation laws if generators of gauge transformations depend on 
derivatives of gauge parameters. At the same time,  dependence of a conserved
flow on gauge parameters make gauge conservation laws form-invariant under
gauge transformations.}
\medskip

\bigskip

\centerline{\bf I. INTRODUCTION}
\bigskip

Energy-momentum in field theory is a vast subject which can be studied from
different viewpoints.  The problem lies in the fact that
the canonical energy-momentum
tensor fails to be a true tensor, while the metric one is appropriate only to
theories with a background geometry. In gravitation theories, an
energy-momentum flow reduces to a superpotential depending on a world vector
field as a gauge parameter.

Analyzing the energy-momentum problem, we follow the general procedure of
obtaining  differential conservation laws in gauge theory [1-6]. Let $Y\to
X$ be a fibre bundle over a world  manifold $X$
coordinated by $(x^\la,y^i)$, where $x^\la$ are coordinates on $X$. By
gauge transformations are generally meant automorphisms of $Y\to
X$ \cite{may}. To obtain differential conservation laws, it
suffices to consider one-parameter groups of gauge transformations whose
generators are projectable vector fields 
\beq
u=u^\m(x^\la)\dr_\m +u^i(x^\la,y^j)\dr_i \label{pr1}
\eeq
on $Y$. If the Lie derivative $\bL_uL$ of a Lagrangian $L$
vanishes, we have the weak conservation law
\beq
d_\la T^\la \ap 0 \label{pr2}
\eeq
of the corresponding symmetry flow $T^\la$ along a vector field $u$. 
If
$u=u^i\dr_i$ is a vertical vector field, (\ref{pr2}) is a familiar N\"other
conservation law, where $T^\la$ is a N\"other current. This is the case of
internal symmetries. If
$u$ is a horizontal prolongation on $Y$ of a vector field $\tau$ on $X$, called
hereafter a world vector field,  we have an energy-momentum conservation law
\cite{sard97,got92,book}. Of course, different horizontal prolongations of
vector fields $\tau$ lead to different energy-momentum flows $T^\la$.

Let us emphasize that the expression (\ref{pr2}) is linear in the vector
field $u$. Therefore, one can consider superposition of conservation laws
along different vector fields. In particular, every vector field
$u$ (\ref{pr1}) on $Y$  projected onto a world vector field $\tau$ can be seen 
as the sum
$u=u_\tau +Êv$ of a horizontal prolongation $u_\tau$ of $\tau$ on $Y$ and a
vertical vector field
$v$ on $Y$. It follows that every conservation law (\ref{pr2}) can be
represented as a superposition of a N\"other conservation law and an
energy-momentum conservation law. Conversely, two horizontal
prolongations $u_\tau$ and $u'_\tau$ of the same world vector field
$\tau$ differ from each other in a vertical vector field
$u'_\tau-u_\tau$. Hence, the energy-momentum flows along $u_\tau$ and
$u'_\tau$ differ from each other in a N\"other current. One
can not single out in a canonical way the N\"other part of an energy-momentum
flow. Therefore, if internal symmetries are broken, an energy-momentum
flow is not conserved in general. 

In particular, a generic gravitational
Lagrangian is invariant under general covariant transformations, but not
under vertical gauge transformations of the general linear group
$GL(4,\bR)$. As a consequence, only the energy-momentum flow, corresponding to
the canonical horizontal prolongation of vector fields
$\tau$  on $X$, i.e., the generators of general covariant transformations, is
generally conserved in gravitation theory. This
 energy-momentum flow possesses the following two important peculiarities. 

(i) It reduces to a superpotential, i.e.,
\beq
T^\la\ap d_\m U^{\m\la}, \qquad U^{\m\la}=-U^{\la\m}. \label{pr3}
\eeq
We will see that this is a common property of gauge conservation laws if a
vector field $u$ depends on the derivatives of gauge parameters. Indeed, the
canonical prolongation $\wt\tau$ of a world vector field $\tau=\tau^\la\dr_\la$
depends on the derivatives $\dr_\m\tau^\la$ of the components of $\tau$ which
play the role of gauge parameters of infinitesimal general covariant
transformations.

(ii) A gravitational superpotential
(\ref{pr3}) depends on the components
$\tau^\la$ of a world vector field $\tau$. This is also a common property of
gauge conservation laws which make them gauge-covariant, i.e., form-invariant
under gauge transformations.  Only in the Abelian case of
electromagnetic theory, an electric current is free from gauge parameters.

A direct computation shows that General Relativity, 
Palatini formalism, metric-affine gravitation theory, and
  gauge gravitation theory in the presence of
fermion fields lead to the same gravitational superpotential.
This is the generalized Komar superpotential
\beq
U^{\m\la}(\tau)= \left(\frac{\dr L}{\dr R_{\m\la}{}^\al{}_\nu}-
\frac{\dr L}{\dr R_{\la\m}{}^\al{}_\nu}\right)(D_\nu\tau^\al +
S_\nu{}^\al{}_\si\tau^\si), \label{K3}
\eeq
where $R_{\m\la}{}^\al{}_\nu$ and $S_\nu{}^\al{}_\si$ are the curvature
and the torsion of a world connection $K_\nu{}^\al{}_\si$, and $D_\nu$ is the
corresponding covariant derivative. In the particular case of the
Hilbert--Einstein Lagrangian $L$ and a Levi--Civita connection $K$, the
superpotential (\ref{K3}) reduces to the familiar Komar superpotential.

Throughout the paper, we follow the convention where the covariant derivative
reads
\be
D_\nu\tau^\al=\dr_\nu\tau^\la -K_\nu{}^\al{}_\m\tau^\m.
\ee
\bigskip

\centerline{\bf II. GAUGE CONSERVATION LAWS}
\bigskip

To obtain gauge conservation laws, we follow the first variational formula
of Lagrangian formalism.

Given a gauge field system described by sections $\f$ of a fibre bundle $Y\to
X$, its space of fields and their first order partial derivatives is
the finite-dimensional first order jet manifold $J^1Y$ of
$Y$. Its elements are fields $\f$ identified by their values and the
values of their first order derivatives at points of
$X$ (see, e.g., \cite{book} for a detailed exposition).  This
space is provided with the adapted coordinates
$(x^\la,y^i,y^i_\la)$, where $y^i_\la$ are the derivative coordinates
such that $y^i_\la\circ\f=\dr_\la\f^i$. The transformation law of the
derivative coordinates is
\be
y'^i_\la=\left(\frac{\dr y'^i}{\dr y^j}y^j_\m +\frac{\dr y'^i}{\dr
x^\m}\right)\frac{\dr x^\m}{\dr x'^\la}.
\ee
A Lagrangian $L$ of a field system on a fibre bundle $Y\to X$ is defined as a
density 
\be
 L=L(x^\la,y^i,y^i_\la)d^4x 
\ee
on the space $J^1Y$. We will use the notation
\be
&& \pi^\la_i=\dr^\la_iL,\\
&& \om=d^4x, \qquad \om_\la=\dr_\la\rfloor\om, \qquad
\om_{\m\la}=\dr_\m\rfloor\om_\la.
\ee

Let $u$ be a projectable vector field (\ref{pr1}) on a fibre bundle $Y$,
treated as the generator of a one-parameter group of gauge transformations.
Its prolongation on the space $J^1Y$ reads
\be
J^1u=u^\la\dr_\la + u^i\dr_i + (d_\la u^i
- y_\m^i\dr_\la u^\m)\dr_i^\la 
\ee
where $d_\la=\dr_\la +y^i_\la\dr_i$ denote the total derivatives. The first
variational formula provides the canonical decomposition of the Lie derivative
\beq
\bL_u L\om=[\dr_\la u^\la L +(u^\la\dr_\la+
u^i\dr_i +(d_\la u^i -y^i_\m\dr_\la u^\m)\dr^\la_i)L]\om \label{04}
\eeq
of a Lagrangian $L$ in accordance with the variational problem. This
decomposition reads
\ben
&& \dr_\la u^\la L +[u^\la\dr_\la+
u^i\dr_i +(d_\la u^i -y^i_\m\dr_\la u^\m)\dr^\la_i]L = \label{bC30'}\\
&& \qquad   (u^i-y^i_\m u^\m )\dl_i L - d_\la T^\la, \nonumber
\een
where 
\beq
\dl_iL=(\dr_i- d_\la\dr^\la_i)L, \qquad d_\la=\dr_\la +y^i_\la\dr_i
+y^i_{\la\m}\dr^\m_i,
\label{b340}
\eeq
are the variational derivatives and 
\beq
T=T^\la\om_\la, \qquad T^\la = \pi^\la_i(u^\m y^i_\m-u^i )-u^\la L, 
 \label{979}
\eeq
is said to be a symmetry flow along the vector field $u$.

It should be emphasized that the flow (\ref{979}) is defined modulo the terms
\be
d_\m(c_i^{\m\la}(y^i_\nu u^\nu-u^i)), 
\ee
where $c_i^{\m\la}$ are arbitrary skew-symmetric functions on $Y$ \cite{book}.
 Here, we leave
aside these boundary terms which  are independent of a Lagrangian, but they
may be essential if one examines integral conservation laws.

On the shell
\be
\delta_i L=(\dr_i- d_\la\dr^\la_i)L=0 
\ee
where solutions of the Euler--Lagrange equations live, the first variational
formula (\ref{bC30'}) leads to the weak  identity 
\ben
&& \dr_\la u^\la L +[u^\la\dr_\la+
u^i\dr_i +(d_\la u^i -y^i_\m\dr_\la u^\m)\dr^\la_i]L \ap \label{b337}\\
&& \qquad - d_\la[\pi^\la_i(u^\m y^i_\m -u^i) -u^\la L]. \nonumber
\een

If the Lie derivative $\bL_uL$
(\ref{04}) vanishes, i.e., the Lagrangian $L$ is invariant under the
corresponding one-parameter group of gauge transformations, the weak identity
(\ref{b337}) is brought into the  weak conservation law
\beq
0\ap - d_\la[\pi^\la_i(u^\m y^i_\m-u^i )-u^\la L], \label{K4'}
\eeq
of the flow $T$ (\ref{979}) along the vector field $u$.

The weak identity (\ref{K4'}) leads to the differential
conservation  law
\be
 d_\la T^\la(\f)=0
\ee
on solutions $\f$ of the Euler--Lagrange equations 
\be
\dl_iL(\f)=0.
\ee
This differential conservation law implies the integral conservation 
law
\beq
\op\int_{\dr M} T^\la (\f)\om_\la =0, \label{b3118}
\eeq
where $M$ is a compact $4$-dimensional sub-manifold of $X$ with a boundary
$\dr M$.

We will see that, in gauge and gravitation models, a conserved flow takes the
special form
\beq
T^\la= W^\la +d_\m U^{\m\la} \label{b381}
\eeq
where the term $W$ is expressed in the variational derivatives (\ref{b340}),
i.e.,
$W\ap 0$,  and $T$ on-shell reduces to a superpotential $U^{\m\la}$
(\ref{pr3}).  In this case, the integral
conservation law (\ref{b3118}) becomes tautological, but the
superpotential form (\ref{b381}) of $T$ implies the following integral
relation
\beq
\op\int_{N^3} T^\la(\f)\om_\la = \op\int_{\dr N^3}
U^{\m\la}(\f) \om_{\m\la}, \label{b3119}
\eeq
where $N^3$ is a compact oriented 3-dimensional submanifold of $X$
with the boundary
$\dr N^3$ and $\f$ is a solution of the Euler--Lagrange equations. One can
think of this relation as being a part of the Euler--Lagrange equations
written in an integral form.
\bigskip

\centerline{\bf III. N\"OTHER CONSERVATIONS LAWS}
\bigskip

Here, we touch on briefly the well-known N\"other conservation laws in gauge
theory of internal symmetries in order to emphasize some common properties of
gauge conservation laws.

Let $P\to X$ be a principal bundle with a structure Lie group $G$. The
corresponding gauge model is formulated on the bundle product 
\beq
Y=E\op\times_XC \label{pr7}
\eeq
 of a
$P$-associated vector bundle $E$ of matter fields and the bundle $C=J^1P/G\to
X$ whose sections $A$ are principal connections on the principal bundle $P$.
The bundle $C$ is equipped with the coordinates $(x^\la,a^m_\la)$ such that
$a^m_\la\circ A=A^m_\la$ are gauge potentials. The bundle $Y$ (\ref{pr7}) is
coordinated by $(x^\la,y^i,a^m_\la)$. Let us consider a one-parameter group of
vertical automorphisms of the principal bundle $P$. They yield the
corresponding one-parameter group of gauge transformations of the product
(\ref{pr7}). Its generator is the vertical vector field 
\be
\xi=(\dr_\mu\x^r+c^r_{qp}a^q_\mu\x^p)\dr^\mu_r +\x^pI_p^i\dr_i,
\ee
where $c^r_{qp}$ are structure constants of the group $G$, $I_p^i$ are
generators of this group on the typical fibre of $Y\to X$, and
$\xi^p(x^\la)$ are gauge parameters, transformed by the coadjoint
representation 
\cite{sard97,book,giach90}.
Let us use the compact notation
\ben
&&\xi = (u^{A\m}_p\dr_\m\x^p+ u^A_p\x^p)\dr_A, \label{C76}\\
&& u_p^{A\m}\dr_A= \dl_p^r\dr^\m_r, \qquad u^A_p\dr_A=
c^r_{qp}a^q_\mu\dr^\mu_r + I_p^i\dr_i. \nonumber
\een 

If a Lagrangian $L$ is gauge invariant, 
the first variational formula
(\ref{bC30'}) leads to the strong equality
\beq
0=(u^A_p\x^p + u^{A\m}_p\dr_\m\x^p)\dl_A L +
d_\la[(u^A_p\x^p + u^{A\m}_p\dr_\m\x^p)\pi^\la_A], \label{b3109}
\eeq
where $\dl_A L$ are the variational derivatives of $L$ and
\be
d_\la =\dr_\la +a^p_{\la\m}\dr_p^\m +y_\la^i\dr_i.
\ee
Due to the arbitrariness of gauge parameters $\x^p(x^\la)$, this equality is
equivalent to the following system of strong equalities:
\bea
&& u^A_p\dl_A L + d_\m(u^A_p\pi^\m_A)=0, \label{D1a}\\
&& u^{A\mu}_p\dl_A L
+ d_\la(u^{A\mu}_p\pi^\la_A) + u^A_p\pi^\mu_A =0,\label{D1b}\\
&& u^{A\la}_p\pi^\mu_A+ u^{A\mu}_p\pi^\la_A=0. \label{D1c}
\eea
Substituting (\ref{D1b}) and (\ref{D1c}) in 
(\ref{D1a}), we obtain the well-known constraint conditions 
\be
u^A_p\dl_A L -d_\m(u^{A\mu}_p\dl_A L)=0
\ee
of the variational
derivatives of a gauge invariant Lagrangian.

On-shell, the first variational formula (\ref{b3109}) leads to the
weak conservation law
\beq
0\ap d_\la[(u^A_p\x^p + u^{A\m}_p\dr_\m\x^p)\pi^\la_A] \label{C300}
\eeq
of the N\"other current
\beq
T^\la=-(u^A_p\x^p + u^{A\m}_p\dr_\m\x^p)\pi^\la_A. \label{b3110}
\eeq
Accordingly, the equalities (\ref{D1a}) -- (\ref{D1c}) on-shell lead to the
equivalent system of N\"other identities 
\bea
&&  d_\m(u^A_p\pi^\m_A)\ap 0, \label{D2a}\\
&& d_\la(u^{A\mu}_p\pi^\la_A)
+ u^A_p\pi^\mu_A \ap 0,\label{D2b}\\
&& u^{A\la}_p\pi^\mu_A+
u^{A\mu}_p\pi^\la_A=0 \label{D2c}
\eea
for a gauge
invariant Lagrangian $L$ \cite{kon}.
They are equivalent to the weak equality (\ref{C300}) due to the
arbitrariness of the gauge parameters $\x^p(x^\la)$. 

The weak identities (\ref{D2a}) -- (\ref{D2c}) play the role of the
necessary and sufficient conditions in order that the weak conservation law
(\ref{C300}) be gauge-covariant. This means
that, if the equality (\ref{C300}) takes place for  gauge parameters
$\x$, it does so for arbitrary deviations $\x + \dl\x$ of $\x$.
Then the conservation law (\ref{C300}) is form-invariant under
gauge transformations, when gauge parameters are transformed by the
coadjoint representation.

The equalities (\ref{D2a}) -- (\ref{D2c}) 
are not independent, for (\ref{D2a}) is a consequence of (\ref{D2b})
and (\ref{D2c}). 
This reflects the fact that, in accordance with the
strong equalities (\ref{D1b}) and
(\ref{D1c}), the N\"other current (\ref{b3110}) is brought 
into the superpotential
form
\beq
T^\la =\x^p
u^{A\la}_p\dl_A L + d_\m U^{\m\la}, \qquad  
U^{\m\la}= -\x^p
u^{A\m}_p\pi^\la_A, \label{pr13}
\eeq
where the superpotential $U^{\m\la}$ does not depend on matter fields.
It is readily observed that the superpotential form of the N\"other current
(\ref{b3110}) is caused by the fact that the vector fields (\ref{C76})
depend on derivatives of gauge parameters.

On solutions of the Euler--Lagrange equations, we have the corresponding
integral relation (\ref{b3119}), which reads
\beq
\op\int_{N^3}T^\la(\f)\om_\la = \op\int_{\dr N^3}
 \x^p \pi^{\m\la}_p(\f) \om_{\m\la}. \label{b3223}
\eeq
 Since the N\"other superpotential (\ref{pr13}) does not contain matter
fields, one can think of (\ref{b3223}) as being the integral relation between
the N\"other current (\ref{b3110}) and the gauge field generated by this
current. Due to the presence of gauge parameters, the relation (\ref{b3223})
is gauge covariant. 

In electromagnetic theory, the similar relation between an
electric current and the electromagnetic field generated by this current is
well known, but it is free from gauge parameters due to the peculiarity of
Abelian gauge models.
Let us consider electromagnetic theory, where 
$G= {\rm U}(1)$ and $I^j(y)=iy^j$.
In this case,
a gauge parameter $\x$ is not changed under gauge transformations.
Therefore, one can put, e.g., $\x=1$. Then the N\"other current (\ref{b3110})
takes the form
\be
T^\la=-u^A\pi^\la_A. 
\ee
Since the group $G$ is Abelian, this current does not depend on
gauge potentials and it is invariant under gauge transformations. We have
\beq
T^\la=-iy^j\pi_j^\la. \label{b3121}
\eeq
 It is easy to
see that $T$ (\ref{b3121}), under the sign change, is the familiar electric
current of matter fields, while the N\"other conservation law (\ref{C300}) is
precisely the equation of continuity. 
The corresponding
integral equation of
continuity (\ref{b3118}) reads
\be
\op\int_{\dr M} (y^j\pi_j^\la)(\f)\om_\la =0. 
\ee
Though the N\"other current $T$ (\ref{b3121}) takes  the superpotential form
\be
T^\la=-\dl^\la L +d_\m U^{\m\la},
\ee
the equation of
continuity is not tautological. This
equation is independent of an electromagnetic field generated by the
electric current (\ref{b3121}) and it is therefore treated as the
strong conservation law of an electric charge. 
When $\x=1$, the electromagnetic superpotential takes the form
\be
U^{\m\la}=-\frac{1}{4\pi}F^{\m\la},
\ee
where $F$ is the electromagnetic strength. 
Accordingly, the integral equality (\ref{b3223}) is the integral form of the
Maxwell equations.  In particular, the well-known relation between the
flux of an electric field through a closed surface and the total electric
charge inside this surface is restated. 
\bigskip

\centerline{\bf IV. ENERGY-MOMENTUM IN GRAVITATION THEORIES}
\bigskip

From now on by a world manifold $X$ is meant a 4-dimensional orientable
noncompact parallelizable manifold. As a consequence, it admits a
pseudo-Riemannian metric and a spin structure. Accordingly, a linear 
connection  and a fibre metric
 on the tangent and cotangent bundles $TX$ and $T^*X$ of $X$ are said
to be a world connection and a world metric, respectively. 

Gravitation theories are formulated on natural bundles $Y\to
X$, e.g., tensor bundles which admit the canonical horizontal prolongations of
any vector field
$\tau$ on
$X$. These prolongations are the generators of general covariant
transformations, where the components of a vector field $\tau$ play the role
of gauge parameters. By the reason we have explained above, we will
investigate the energy-momentum conservation laws associated with these
prolongations.
\bigskip

\centerline{\bf A. Tensor fields}
\bigskip

We start from tensor fields which clearly illustrate the main
peculiarities of energy-momentum conservation laws on natural bundles.
Let 
\beq
Y=(\op\ot^mTX)\op\ot_X(\op\ot^kT^*X) \label{971}
\eeq
be a tensor bundle equipped with the holonomic coordinates $(x^\la,
\dot x^{\al_1\cdots\al_m}_{\bt_1\cdots\bt_k})$.

Given a vector field $\tau=\tau^\la\dr_\la$, we have its
canonical prolongation 
\beq
\wt\tau = \tau^\m\dr_\m + [\dr_\nu\tau^{\al_1}\dot
x^{\nu\al_2\cdots\al_m}_{\bt_1\cdots\bt_k} + \ldots
-\dr_{\bt_1}\tau^\nu \dot x^{\al_1\cdots\al_m}_{\nu\bt_2\cdots\bt_k}
-\ldots]\frac{\dr}{\dr \dot
x^{\al_1\cdots\al_m}_{\bt_1\cdots\bt_k}} \label{l28}
\eeq
on the tensor bundle (\ref{971}) and, in particular, its prolongations
\beq
\wt\tau = \tau^\m\dr_\m +\dr_\nu\tau^\al\dot x^\nu\frac{\dr}{\dr\dot x^\al}
\label{l27}
\eeq
on the tangent bundle $TX$ and
\be
\wt\tau = \tau^\m\dr_\m -\dr_\bt\tau^\nu\dot x_\nu\frac{\dr}{\dr\dot x_\bt}
\ee
on the cotangent bundle $T^*X$. 

Of course, one can consider the horizontal prolongation 
\beq
\tau_K =\tau^\la(\dr_\la +K_\la{}^\bt{}_\al\dot x^\al\frac{\dr}{\dr\dot
x^\bt}) \label{b3180}
\eeq
of a world vector field
$\tau$ on $TX$ and tensor bundles 
by means of any world connection $K$. This is the generator of a
1-parameter group of nonholonomic automorphisms of
these bundles. These automorphisms are met with in 
gauge theory of the general linear group $GL(4,\bR)$ \cite{heh}, but a
generic gravitational Lagrangian is not invariant under these transformations.
Note that the prolongations (\ref{l27}) and (\ref{b3180}) were treated as
generators of the gauge group of translations in the pioneer gauge gravitation
models (see
\cite{heh76,iva} for a survey).

Let 
us use the compact notation 
$y^A = \dot x^{\al_1\cdots\al_m}_{\bt_1\cdots\bt_k}$
such that the canonical prolongation $\wt\tau$ (\ref{l28}) reads 
\beq
\wt\tau =\tau^\la\dr_\la + u^A{}_\al^\bt\dr_\bt\tau^\al\dr_A. \label{Q37}
\eeq
This expression  is the general form of the canonical prolongation
of a world vector field $\tau$ on a natural bundle $Y$,
when this prolongation depends only on the first order partial derivatives of
the components of
$\tau$. Therefore, the results obtained below for tensor fields are also true
 for every such natural bundle $Y$.

Let a Lagrangian $L$ of tensor fields 
be invariant under general covariant transformations, i.e., its Lie derivative
(\ref{04}) vanishes for any world vector field $\tau$. Then we have the strong
equality
\beq
\dr_\al(\tau^\al L) + u^A{}_\al^\bt\dr_\bt\tau^\al\dr_A L +
d_\m(u^A{}_\al^\bt\dr_\bt\tau^\al)\pi_A^\m -
 y^A_\al\dr_\bt\tau^\al\pi_A^\bt =0. \label{C310}
\eeq
The corresponding weak identity (\ref{K4'}) takes the form
\beq
0\ap - d_\la [\pi^\la_A(y^A_\al\tau^\al-u^A{}_\al^\bt\dr_\bt\tau^\al)
-\tau^\la L]. \label{Q36}
\eeq
Due to the arbitrariness of the gauge parameters $\tau^\al$, the equality
(\ref{C310}) is equivalent to the system of strong equalities
\bea
&& \dr_\la L=0, \\
&& \dl^\bt_\al L + u^A{}_\al^\bt\dl_A L + d_\m(u^A{}_\al^\bt\pi_A^\m) =
 y^A_\al\pi_A^\bt,\label{C311a}\\
&& u^A{}_\al^\bt\pi_A^\m +u^A{}_\al^\m\pi_A^\bt =0, \label{C311b}
\eea
where  $\dl_A L$ are the variational derivatives.

Substituting the relations (\ref{C311a}) and (\ref{C311b})
in (\ref{Q36}), we obtain the energy-momentum conservation law
\beq
0\ap - d_\la [u^A{}_\al^\la\dl_A L\tau^\al +
 d_\m(u^A{}_\al^\la\pi_A^\m\tau^\al)].
\label{C313}
\eeq 
A glance at this expression shows that, on-shell, the
corresponding energy-momentum flow reduces to a superpotential, i.e.,
\beq
T^\la= \tau^\al u^A{}_\al^\la\dl_A L +
 d_\m U^{\m\la}, 
\qquad U^{\m\la} = u^A{}_\al^\la\pi_A^\m\tau^\al. \label{972}
\eeq
It is readily seen that the energy-momentum superpotential (\ref{972}) emerges
from the dependence of the canonical prolongation $\wt\tau$ (\ref{Q37}) on the
derivatives of the components of the vector field $\tau$.
This dependence guarantees that
the energy-momentum conservation law (\ref{C313}) is maintained under general
covariant transformations.
\bigskip

\centerline{\bf B. General Relativity}
\bigskip

A pseudo-Riemannian metric $g$ on a world manifold $X$ is represented by a
section of the quotient
\beq
\Si=LX/O(1,3)\to X, \label{pr15}
\eeq
where $LX$ is the bundle of linear frames in the tangent
spaces to $X$. It is called the metric bundle. The linear frame bundle $LX$
is a principal bundle with the structure group $GL(4,\bR)$. For the sake of
simplicity, we will identify the metric bundle with a sub-bundle of the tensor
bundle
\beq
\Si\subset \op\vee^2TX, \label{pr10}
\eeq
coordinated by $(x^\la,g^{\m\nu})$. In General Relativity,
it
is more convenient to consider the metric bundle as a sub-bundle of the
tensor bundle
\beq
\Si\subset\op\vee^2T^*X \label{pr11}
\eeq
equipped with the coordinates $(x^\la,g_{\m\nu})$. 

The second order Hilbert--Enstein Lagrangian $L_{HE}$ of General Relativity is
defined on the second order jet manifold
$J^2\Si$ of $\Si$ coordinated by
$(x^\la, g_{\al\bt},\, g_{\la\al\bt},\, g_{\m\la\al\bt}).$ It reads
\beq
L_{HE}=-\frac{1}{2\kp}g^{\al\nu}g^{\bt\mu}R_{\m\nu\al\bt}
\sqrt{ -g}\om, \qquad g={\rm det}(g_{\m\nu}),\label{C127}
\eeq
where
\be
&& R_{\m\nu\al\bt} = - [\frac12 (g_{\m\bt\al\nu}+ g_{\nu\al\bt\m}
-g_{\nu\bt\al\m} -g_{\m\al\bt\nu}) +  \\
&& \qquad g_{\ve\g}(\{_\m{}^\ve{}_\bt\}\{_\nu{}^\g{}_\al\} -
\{_\nu{}^\ve{}_\bt\}\{_\m{}^\g{}_\al\})],  \\
&& \{_\nu{}^\al{}_\m\} = -\frac12g^{\al\bt}(g_{\nu\bt\m} +g_{\m\bt\nu}
- g_{\bt\m\n}). 
\ee
Let $\tau$ be a vector field on $X$ and 
\beq
\wt\tau =\tau^\la\dr_\la - (g_{\nu\bt}\dr_\al\tau^\nu
+g_{\al\nu}\dr_\bt\tau^\nu)\dr^{\al\bt} \label{b3162}
\eeq
its canonical prolongation (\ref{l28}) on the metric bundle $\Si$
(\ref{pr11}). Since the Lagrangian
 $L_{HE}$ (\ref{C127}) is invariant under  general covariant
transformations, its Lie derivative along the vector field $\wt\tau$
(\ref{b3162}) vanishes.  Then the first variational formula for second order
Lagrangians \cite{book,nov}
 leads to the energy-momentum conservation law  
\be
&& 0 \ap -d_\la\{2g_{\m\al}\tau^\m\dl^{\al\la} L_{HE} + 
d_\m[\frac1{2\kp}\sqrt{-g}(g^{\la\nu}\nabla_\nu\tau^\m -
g^{\m\nu}\nabla_\nu\tau^\la)]\}, \\
&& \nabla_\nu\tau^\m=\dr_\nu\tau^\m-\{_\nu{}^\m{}_\al\}\tau^\al.
\ee
A glance at this conservation law shows that, on-shell, the energy-momentum
flow reduces to the well-known Komar superpotential \cite{kom}:
\beq
U^{\m\la} = \frac1{2\kp}\sqrt{\nm g}(g^{\la\nu}
\nabla_\nu\tau^\m -
g^{\m\nu}\nabla_\nu\tau^\la). \label{b3163}
\eeq
\bigskip

\centerline{\bf C. Metric-affine gravitation theory}
\bigskip

In metric-affine gravitation theory, gravity is described by
a pseudo-Riemannian metric $g$ and a world connection $K$ on $X$.
Since world connections are associated with  principal connections on
the linear frame bundle $LX$, there is one-to-one correspondence
between the world connections and the sections of 
the quotient  
\beq
C_K=J^1LX/GL(4,\bR)\to X. \label{015}
\eeq
This bundle is provided with the coordinates $(x^\la, k_\la{}^\nu{}_\al)$ so
that, for any section
$K$ of $C_K\to X$,
\be
k_\la{}^\nu{}_\al\circ K=K_\la{}^\nu{}_\al
\ee
are the coefficients of the world connection $K$. 
The bundle $C_K$  (\ref{015}) admits the
canonical horizontal prolongation 
\beq
\wt\tau_K = \tau^\m\dr_\m +[\dr_\nu\tau^\al k_\m{}^\nu{}_\bt -
\dr_\bt\tau^\nu k_\m{}^\al{}_\nu - \dr_\m\tau^\nu
k_\nu{}^\al{}_\bt + \dr_{\m\bt}\tau^\al]\frac{\dr}{\dr k_\m{}^\al{}_\bt}
\label{b3150}
\eeq
of vector fields $\tau$ on $X$.
We will use the compact notation
\be
\wt\tau_K =\tau^\la\dr_\la + (u^A{}_\g^\ve\dr_\ve\tau^\g
+u^A{}_\g^{\ve\si}\dr_{\ve\si}\tau^\g)\dr_A, 
\ee
 where
\ben
&& y^A=k_\m{}^\al{}_\bt, \label{pr30} \\ 
&&  u_\m{}^\al{}_\bt{}^\ve_\g= k_\m{}^\ve{}_\bt
\dl^\al_\g -k_\m{}^\al{}_\g
\dl^\ve_\bt - k_\g{}^\al{}_\bt \dl^\ve_\m, \nonumber\\
&&  u_\m{}^\al{}_\bt{}^{\ve\si}_\g = \dl^\ve_\m
\dl^\si_\bt \dl^\al_\g. \nonumber
\een

Metric-affine gravitation theory is formulated on the bundle product
\beq
Y=\Si\op\times_XC_K, \label{pr12}
\eeq
coordinated by $(x^\la, g^{\al\bt}, k_\m{}^\al{}_\bt)$, where $\Si$ is the
sub-bundle (\ref{pr10}).  The corresponding  space $J^1Y$ is
equipped with the coordinates  
\be
(x^\la, g^{\al\bt}, k_\m{}^\al{}_\bt,  g_\la{}^{\al\bt}, 
k_{\la\m}{}^\al{}_\bt).
\ee
We will assume that a metric-affine Lagrangian $L_{MA}$ factorizes through
the curvature 
\be
R_{\la\m}{}^\al{}_\bt = k_{\la\m}{}^\al{}_\bt - k_{\m\la}{}^\al{}_\bt
+ k_\m{}^\al{}_\ve k_\la{}^\ve{}_\bt
-k_\la{}^\al{}_\ve k_\m{}^\ve{}_\bt, 
\ee
 and does not depend on the
derivative coordinates $g_\la{}^{\al\bt}$ of a world metric. Then the
following relations take place:
\ben
&&  \pi^{\la\nu}{}_\al{}^\bt= -\pi^{\nu\la}{}_\al{}^\bt, \label{K300'}\\
&&\frac{\dr L_{MA}}{\dr k_\nu{}^\al{}_\bt}= 
\pi^{\la\nu}{}_\al{}^\si k_\la{}^\bt{}_\si
-\pi^{\la\nu}{}_\si{}^\bt k_\la{}^\si{}_\al. \label{K300}
\een 
We also have the equalities
\be
\pi^\la_A u^A{}_\al^{\bt\m} =\pi^{\la\m}{}_\al{}^\bt,\qquad \pi^\ve_A
u^A{}_\al^\bt = -\dr^\ve{}_\al{}^\bt L_{\rm MA} -
\pi^{\ve\bt}{}_\si{}^\g k_\al{}^\si{}_\g. 
\ee

Given a vector field $\tau$ on a world manifold $X$, its canonical
prolongation on the product (\ref{pr12}) reads
\be
&&\wt\tau =\tau^\la\dr_\la +
(g^{\nu\bt}\dr_\nu\tau^\al +g^{\al\nu}\dr_\nu\tau^\bt)\dr_{\al\bt}+\\ 
 && \qquad (u^A{}_\al^\bt\dr_\bt\tau^\al
+u^A{}_\al^{\bt\m}\dr_{\bt\m}\tau^\al)\dr_A.
\ee
Let a metric-affine Lagrangian $L_{MA}$ be invariant under
general covariant transformations, i.e.,
\beq
\bL_{\wt\tau}L_{MA}=0 \label{b3172}
\eeq
for any world vector field $\tau$.
Then, on-shell, the first variational formula (\ref{bC30'}) leads to
the weak conservation law
\beq
0\ap - d_\la[ \pi^\la_A(y^A_\al\tau^\al -u^A{}_\al^\bt\dr_\bt\tau^\al
 -u^A{}_\al^{\ve\bt}\dr_{\ve\bt}\tau^\al) -\tau^\la L_{MA}] \label{K8}\\
\eeq
where
\beq
T^\la= \pi^\la_A(y^A_\al\tau^\al -u^A{}_\al^\bt\dr_\bt\tau^\al
 -u^A{}_\al^{\ve\bt}\dr_{\ve\bt}\tau^\al) -\tau^\la L_{MA} \label{b3190}
\eeq
is the energy-momentum flow of the metric-affine gravity.

It is readily observed that, in the local gauge where the vector
field
$\tau$ is constant, the energy-momentum flow (\ref{b3190}) 
leads to the canonical energy-momentum tensor
\be
T^\la =(\pi^{\la\m}{}_\bt{}^\nu k_{\al\m}{}^\bt{}_\nu
-\dl^\la_\al L_{\rm MA})\tau^\al.
\ee
This tensor was suggested in order to describe the energy-momentum complex in
the Palatini model \cite{mur}.

Due to the arbitrariness of the gauge parameters $\tau^\la$, the equality
(\ref{b3172}) is equivalent to the system of strong equalities
\ben
&& \dr_\la L=0, \nonumber \\
&& \dl^\bt_\al L_{MA} + 2g^{\bt\m}\dl_{\al\m} L_{MA} + u^A{}_\al^\bt\dl_A
L_{MA} + d_\m(\pi^\m_A  u^A{}_\al^\bt)
- y^A_\al\pi^\bt_A=0,
\label{b3173b} \\ 
&& (u^A{}_\g^{\ve\si}\dr_A + u^A{}_\g^\ve\dr^\si_A) L_{\rm MA}  
\dr_{\si\ve}\tau^\g= 0, \label{b3173c}\\
&& \pi^{(\la\ve}{}_\g{}^{\si)}=0, \label{b3173d}
\een
where $\dl_{\al\m} L_{MA}$ and $\dl_A L_{MA}$ are the
corresponding variational derivatives.
It is readily observed that 
the equality (\ref{b3173c}) holds owing to the relation (\ref{K300}), while 
the equality (\ref{b3173d}) does due to the relation (\ref{K300'}).

Substituting the term 
$y^A_\al\pi^\bt_A$
 from the
expression (\ref{b3173b}) in the energy-momentum conservation law (\ref{K8}),
we bring  this conservation law into the form
\ben
&& 0\ap -
d_\la[2g^{\la\m}\tau^\al\dl_{\al\m} L_{MA} + u^A{}_\al^\la\tau^\al\dl_A
L_{MA} 
 - \pi^\la_Au^A{}_\al^\bt\dr_\bt\tau^\al + \label{b3174}\\
&& \qquad d_\m(\pi^{\la\m}{}_\al{}^\bt)
\dr_\bt\tau^\al +
d_\m(\pi^\m_A  u^A{}_\al^\la)\tau^\al - d_\m(\pi^{\la\m}{}_\al{}^\bt
\dr_\bt\tau^\al) ].
\nonumber
\een
After separating the variational derivatives, the energy-momentum
conservation law (\ref{b3174}) of the metric-affine gravity takes the
superpotential form 
\be
&& 0\ap - d_\la [2g^{\la\m}\tau^\al\dl_{\al\m} L_{MA}
+(k_\m{}^\la{}_\g\dl^\m{}_\al{}^\g L_{MA} -
 k_\m{}^\si{}_\al\dl^\m{}_\si{}^\la L_{MA} -
k_\al{}^\si{}_\g\dl^\la{}_\si{}^\g L_{MA})\tau^\al +  \\
&& \qquad \dl^\la{}_\al{}^\m L_{MA}\dr_\m\tau^\al
-d_\m(\dl^\m{}_\al{}^\la L_{MA})\tau^\al + 
 d_\m(\pi^{\m\la}{}_\al{}^\nu(\dr_\nu\tau^\al
-k_\si{}^\al{}_\nu\tau^\si))],
\ee
where the energy-momentum flow on-shell reduces to the  generalized Komar 
superpotential
\beq
U^{\m\la}= \frac{\dr L_{MA}}{k_{\m\la}{}^\al{}_\nu}(\dr_\nu\tau^\al
-k_\si{}^\al{}_\nu\tau^\si) \label{K3'}
\eeq
 that we have written in the form (\ref{K3}) \cite{giachcqg}.

In particular, let us consider the Hilbert--Einstein Lagrangian density
\be
 L_{\rm HE}=-\frac{1}{2\kp}R\sqrt{-g}\om,\qquad 
 R=g^{\la\nu}R_{\la\al}{}^\al{}_\nu, 
\ee
in the framework of metric-affine gravitation theory. 
Then
the generalized Komar superpotential (\ref{K3}) comes to the Komar
superpotential (\ref{b3163}) if we substitute the Levi--Civita connection
$k_\nu{}^\al{}_\si =\{_\nu{}^\al{}_\si\}$.
One may generalize this example by considering the Lagrangian 
\be
L= f(R)\sqrt{-g}\om,
\ee
where $f(R)$ is a polynomial of the scalar curvature $R$.
In the case of a symmetric connection, we restate the superpotential
\be
U^{\m\la}=-\frac{\dr f}{\dr R}\sqrt{-g}
(g^{\la\nu}D_\nu\tau^\m - g^{\m\nu}D_\nu\tau^\la)
\ee
of the Palatini model \cite{bor} as like as the superpotential in the recent 
work \cite{bor98} where Lagrangians of the Palatini model factorize through
the product $R^{\al\bt}R_{\al\bt}$.
\bigskip

\centerline{\bf D. Gauge gravitation theory}
\bigskip

Turning to the energy-momentum conservation law in gauge gravitation theory, we
meet with the problem that spinor bundles over a world manifold do not
admit general covariant transformations. This difficulty can be overcome as
follows
\cite{book,sard98}. The linear frame bundle $LX$ is the principal
bundle
$LX\to\Si$ over the metric bundle $\Si$ (\ref{pr15}) whose structure group is
the Lorentz group. Let us consider a spinor bundle $S\to\Si$ associated with
the Lorentz bundle $LX\to\Si$. For each gravitational field $g$, the
restriction of this spinor bundle to $g(X)\subset\Si$ is isomorphic to the
spinor bundle $S^g\to X$ whose sections describe Dirac fermion
fields in the presence of a gravitational field $g$. 
Moreover, the spinor bundle $S\to\Si$ can be provided with the Dirac operator
and the Dirac Lagrangian whose restrictions to $g(X)$ are the familiar Dirac
operator and Dirac Lagrangian of fermion fields in the presence of a background
gravitational field $g$. It follows that sections of the fibre bundle
$S\to X$ describe the total system of fermion and gravitational fields on a
world manifold $X$. At the same time, the fibre bundle
$S\to X$ is not a spinor bundle, and it inherits general covariant
transformations of the frame bundle
$LX$. The corresponding horizontal prolongation on $S$ of world vector fields
on
$X$ can be constructed.

As a consequence, gauge gravitation theory of metric gravitational fields,
world connections and Dirac fermion fields can be formulated on the the bundle
product
\beq
 Y=C_K \op\times_\Si S, \label{042}
\eeq
coordinated by $(x^\m,h^\m_a, k_\m{}^\al{}_\bt,\psi^A)$, where $h^\m_a$ are
tetrad coordinates on $\Si$, i.e., 
\be
g^{\m\nu}=h^\m_a h^\nu_b\eta^{ab}
\ee
where $\eta$ is the Minkowski metric. The corresponding
 space $J^1Y$ is provided with the adapted coordinates 
\be
(x^\m,h^\m_a,
k_\m{}^\al{}_\bt,\psi^A,h^\m_{\la a}, k_{\la\m}{}^\al{}_\bt,\psi^A_\la).
\ee
The total Lagrangian
on this  space is the sum 
\beq
L=L_{MA} + L_D\label{y12}
\eeq
of the metric-affine Lagrangian $L_{MA}$ in the previous Section and the
Dirac Lagrangian
\be
&& L_D=\{\frac{i}{2}h^\la_q[\psi^+_A(\g^0\g^q)^A{}_B(\psi^B_\la-
\frac14(\eta^{kb}h^a_\m
-\eta^{ka}h^b_\m)(h^\m_{\la k} -h^\nu_k
k_\la{}^\m{}_\nu)L_{ab}{}^B{}_C\psi^C)- \\
&& \qquad (\psi^+_{\la A} -
\frac14(\eta^{kb}h^a_\m
-\eta^{ka}h^b_\m)(h^\m_{\la k} -h^\nu_k
k_\la{}^\m{}_\nu)\psi^+_C L^+_{ab}{}^C{}_A)(\g^0\g^q)^A{}_B\psi^B]-
\label{b3265}\\  &&\qquad  m\psi^+_A(\g^0)^A{}_B\psi^B\}\sqrt{-g}, \qquad
L_{ab}=\frac14[\g_a,\g_b].
\ee
Note that, in fact, the Dirac Lagrangian $L_D$ depends only
on the torsion
$k_\la{}^\m{}_\nu -k_\nu{}^\m{}_\la$ of a world connection, while the
pseudo-Riemannian part is given by the derivative coordinates $h^\m_{\la
k}$.

Given a vector field $\tau$ on a world manifold $X$, its horizontal
prolongation on the product (\ref{042}) is
\ben
&& \wt\tau_Y= \wt\tau + v, \label{pr20}\\
&& \wt\tau = \wt\tau_K + \dr_\nu\tau^\m h^\nu_c 
\frac{\dr}{\dr h^\m_c}, \nonumber \\
&& v= \frac14  Q^\m_k(\eta^{kb}h^a_\m-\eta^{ka}h^b_\m)
[-L_{ab}{}^d{}_ch^\nu_d\frac{\dr}{\dr h^\nu_c} +
L_{ab}{}^A{}_B\psi^B\dr_A + L^+_{ab}{}^A{}_B\psi^+_A\dr^B], \nonumber
\een
where $\wt\tau_K$ is the vector field (\ref{b3150}),
 $L_{ab}{}^d{}_c$ are generators of the Lorentz group in the 
Minkowski space, and the terms $Q^\m_c$ obey the condition
\be
(Q^\m_a h^\nu_b + Q^\nu_a h^\m_b)\eta^{ab}= 0
\ee
(see \cite{book,sard98} for a detailed exposition). The horizontal part
$\wt\tau$ of the vector field (\ref{pr20}) is the generator of a one-parameter
group of general covariant transformations of the fibre bundle (\ref{042}),
whereas the vertical one $v$ is the generator of a one-parameter group of
vertical Lorentz automorphisms of the spinor bundle $S\to\Si$. By
construction, the total  Lagrangian $L$ (\ref{y12}) obeys the relations
\ben
&& \bL_v L_D=0. \label{K200'}\\
&& \bL_{\wt\tau}L_{MA}=0, \qquad \bL_{\wt\tau}L_D=0.\label{K200}
\een 
The relation (\ref{K200'}) results in the N\"other conservation law,
while the equalities (\ref{K200}) lead to the energy-momentum one
\cite{giach97,sard97b}.

Using the compact notation (\ref{pr30}), let us rewrite the horizontal part
$\wt\tau$ of the vector field (\ref{pr20}) in the form 
\be
\wt\tau = \tau^\m\dr_\m + \dr_\nu\tau^\m h^\nu_a\frac{\dr}{\dr h^\m_a} +
(u^A{}_\al^\bt\dr_\bt\tau^\al + u^A{}_\al^{\ve\bt}\dr_{\ve\bt}\tau^\al)\dr_A.
\ee
 Due to the arbitrariness of the functions
$\tau^\al$, the conditions (\ref{K200}) lead to the strong equalities 
\ben
&& \dl^\bt_\al L_{MA} + 2h^{\bt\m}\dl_{\al\m}L_{MA} +
u^A{}_\al^\bt\dl_AL_{MA}  + d_\m(\pi^\m_A  u^A{}_\al^\bt)
=y^A_\al\dr_A^\bt L_{MA},
\label{K9} \\
&&\dl_\al^\bt L_{D} +\sqrt{-g} t_\al^\bt +
\frac{\dr L_{D}}{\dr h^\al_{\la c}}
h^\bt_{\la c} + \dr_AL_{D} u^{A\bt}_\al= \label{K301}\\
&& \qquad \frac{\dr L_{D}}{\dr h^\m_{\bt c}}h^\m_{\al c} +
 \frac{\dr L_{D}}{\dr \psi^A_\bt} \psi^A_\al +
\frac{\dr L_{D}}{\dr \psi^+_{\bt A}}\psi^+_{\al A}, \nonumber 
\een 
where
\be
\sqrt{-g} t^\bt_\al =h^\bt_a\frac{\dr L_{D}}{\dr h^\al_a}
\ee
is the metric energy-momentum tensor of fermion fields.
We also have the relations (\ref{K300'}), (\ref{K300}) and 
\beq
\frac{\dr L_{D}}{\dr k_\la{}^\m{}_\nu} =-\frac{\dr L_{D}}{\dr
k_\nu{}^\m{}_\la}=\frac{\dr L_{D}}{\dr h^\m_{\la c}} h^\nu_c. \label{b3.5000}
\eeq
The corresponding energy-momentum conservation law reads
\ben
&&0\ap - d_\la[ \dr^\la_A L_{\rm
MA}(y^A_\al\tau^\al - u^A{}_\al^\bt\dr_\bt\tau^\al
-u^A{}_\al^{\ve\bt}\dr_{\ve\bt}\tau^\al) - \label{K400}\\ 
&& \qquad
\frac{\dr L_{D}}{\dr h^\al_{\la c}} (\dr_\bt\tau^\al h^\bt_c -
h^\al_{\m c}\tau^\m) + \frac{\dr L_{D}}{\dr \psi^A_\la} \psi^A_\al\tau^\al
+
\frac{\dr L_{D}}{\dr \psi^+_{\la A}} \psi^+_{\al A}\tau^\al
-\tau^\la L]. \nonumber
\een

Substituting the term $y^A_\al\dr_A^\bt L_{MA}$  from 
(\ref{K9}) and the term
\be
\frac{\dr L_{D}}{\dr h^\m_{\bt c}}h^\m_{\al c} +
 \frac{\dr L_{D}}{\dr \psi^A_\bt} \psi^A_\al +
\frac{\dr L_{D}}{\dr \psi^+_{\bt A}}\psi^+_{\al A}
\ee
 from (\ref{K301}) 
in (\ref{K400}), we bring this conservation law into the superpotential form
\ben
&& 0\ap - d_\la [2h^{\la\m}\tau^\al\dl_{\al\m} L
+(k_\m{}^\la{}_\g\dl^\m{}_\al{}^\g L -
 k_\m{}^\si{}_\al\dl^\m{}_\si{}^\la L -
k_\al{}^\si{}_\g\dl^\la{}_\si{}^\g L)\tau^\al +  \label{K303} \\
&& \qquad \dl^\la{}_\al{}^\m L\dr_\m\tau^\al
-d_\m(\dl^\m{}_\al{}^\la L)\tau^\al + 
 d_\m(\pi^{\m\la}{}_\al{}^\nu(\dr_\nu\tau^\al
-k_\si{}^\al{}_\nu\tau^\si))]  -\nonumber\\
&&\qquad  d_\la[(\frac{\dr L_{D}}{\dr h^\al_{\m a}}h^\la_a +
\frac{\dr L_{D}}{\dr h^\al_{\la a}}h^\m_a)\dr_\m\tau^\al]. \nonumber
 \een
By virtue of the relations (\ref{b3.5000}), the last term in the expression 
(\ref{K303}) vanishes, i.e., fermion fields do not
 contribute to the superpotential. It follows that the energy-momentum
conservation law (\ref{K400}) of gauge gravitation theory takes the
superpotential form (\ref{b381}),  where
$U^{\m\la}$ is the generalized  Komar superpotential (\ref{K3'}). 
\bigskip

\centerline{\bf V. CONCLUSIONS}
\bigskip

The above energy-momentum
conservation laws in gravitation theories are derived from
the condition of invariance of gravitational Lagrangians
under general covariant transformations. We have shown that they possess the
generic properties of gauge conservation laws. Since the generators of
general covariant transformations depend on the derivatives of gauge
parameters, i.e., components of a world vector field $\tau$, the
corresponding energy-momentum flow $T^\la$ takes the superpotential form.
As in electromagnetic theory, one can write the integral relation
(\ref{b3119}) on-shell between the energy-momentum flow $T^\la$ and the
gravitational superpotential $U^{\m\la}$. The examples of fermion fields and
Proca fields
\cite{book} show that the superpotential
$U^{\m\la}$ does not depend on matter fields. Then, by analogy with
electromagnetic and gauge theories, one can think of (\ref{b3119}) as being 
the relation between the energy-momentum flow $T^\la$ in a compact
3-dimensional manifold $N\subset X$ and the flux $U^{\m\la}$ of a
gravitational field generated by this flow through the boundary $\dr N$. 

The peculiarity of
gravitation theory as like as any non-Abelian gauge theory lies in the fact
that the integral relation (\ref{b3223}) depends on a gauge parameter. In the
case of general covariant transformations, this gauge parameter is a world
vector field $\tau$ whose canonical prolongation on a natural bundle is the
generator of general covariant transformations. As a consequence, the
energy-momentum conservation law in gravitation theories as like as
in mechanics \cite{98} depends on a reference frame, but it is
 form-invariant under general covariant
transformations. In particular, with respect to the local gauge
$\tau=$const., the energy-momentum flow in the conservation law (\ref{K400})
is the canonical energy-momentum tensor.

A direct computation also shows that, for a number of gravitation models, the
generalized Komar superpotential (\ref{K3}) appears to be an 
energy-momentum superpotential associated with gauge invariance under general
covariant transformations.

\end{document}